\documentclass[prb,twocolumn,showpacs,amsmath,amssymb]{revtex4}

\usepackage{graphicx}
\usepackage{dcolumn}
\usepackage{bm}

\begin{document}


\title{Nontrivial in-plane-magnetic-field dependence of THz wave emission\\
from intrinsic Josephson junctions controlled by surface impedance}

\author{Yoshihiko Nonomura}
 \email{nonomura.yoshihiko@nims.go.jp}
\affiliation{%
Computational Materials Science Unit, National Institute for Materials Science, 
Tsukuba, Ibaraki 305-0047, Japan
}%

\date{\today}

\begin{abstract}
In THz wave emission from intrinsic Josephson junctions in in-plane 
magnetic fields, emission intensity strongly depends on the surface 
impedance $Z$ similarly to the case without external magnetic fields. 
Cavity resonance modes are stabilized for $Z \ge 3$, and the fundamental 
mode gives the strongest emission. As the in-plane magnetic field increases 
for a fixed number of junctions, dynamical phase transitions seem to occur 
between the $\pi$-phase-kink state, various incommensurate phase-kink 
states, and in-phase state. As $Z$ varies, a crossover of the field 
profile of maximum intensity takes place for $Z \approx 50$ between 
characteristic peaks for smaller $Z$ (typically $Z \approx 30$) 
and monotonic decrease for larger $Z$ (typically $Z \approx 70$). 
The double-peak structure reported in experiments can be explained 
for $Z=30$ by finite-size analysis with respect to number of junctions.
\end{abstract}
\maketitle

{\it Introduction.}
Although THz wave emission from intrinsic Josephson junctions 
(IJJs) had been investigated in in-plane magnetic fields, 
experimental realization of such emission~\cite{Bae} 
had been quite difficult. Evident emission was observed without 
external magnetic fields,~\cite{Ozyuzer,Kadowaki} where IJJs have 
about a thousand of junctions so that the velocity of Josephson 
plasma mode is already as fast as that of light in IJJs.~\cite{Sakai}  
Then, possible emission states were investigated theoretically, and the 
uniform in-phase state \cite{Matsumoto,Koyama} and the $\pi$-phase-kink 
states with symmetry breaking along the $c$ axis \cite{Lin08,Koshelev} 
have been proposed. The present author showed \cite{Nonomura09} 
that these states are both stationary according to the bias current $J$ 
and surface impedance $Z$. Recently emission in in-plane fields has 
been investigated again. In an experiment intensity decreases 
monotonically as the in-plane field increases, \cite{Welp} while 
in another experiment the emission intensity seems to have 
some characteristic peaks in the field profile. \cite{Yamaki} 
The present study suggests that these two experiments 
may not be contradictory.
\medskip
\par
{\it Model and formulation.}
As long as thermal fluctuations are not taken into account in the 
modeling, dimensional reduction along the in-plane magnetic field 
is justified. When the direction of the magnetic field is chosen as 
the $y$ axis, the basic equations are given by~\cite{Tachiki}
\begin{eqnarray}
\label{eq-1}
\partial_{x'}^{2}\psi_{l}
&=&(1-\zeta\Delta^{(2)})\left(\partial_{t'}E'_{l}+\beta E'_{l}+\sin \psi_{l}-J'\right),\\
\partial_{t'}\psi_{l}
&=&(1-\alpha\Delta^{(2)})E'_{l},
\label{eq-2}
\end{eqnarray}
where $\Delta^{(2)}$ is defined in 
$\Delta^{(2)}X_{l} \equiv X_{l+1}-2X_{l}+X_{l-1}$, and  
explicit expressions of scaled quantities in the insulating layers 
are given in Eqs.\ (7)--(10) of Ref.~9; e.g.\ length is scaled by 
the penetration depth $\lambda_{c}$ in $x'$, time by inverse 
of the plasma frequency $\omega_{\rm p}$ in $t'$, and bias 
current by the critical current $J_{\rm c}$ in $J'$. 
In addition to the scaled electric field $E'_{l}$, the scaled magnetic field 
$B'_{l}$ is obtained from $\partial_{x'} \psi_{l}=(1-\zeta\Delta^{(2)})B'_{l}$. 
Using the material parameters of Bi$_{2}$Sr$_{2}$CaCu$_{2}$O$_{8}$ 
in Ref.~12, we have a large inductive coupling $\zeta=4.4\times 10^{5}$ 
and a small capacitive coupling $\alpha=0.1$, and 
a scaled conductivity $\beta=0.02$ is taken.

Although a thousand of junctions are essential for obtaining 
the plasma velocity as fast as that of light in IJJs, 
it is almost impossible to arrive at stationary states for such a 
large number of junctions numerically at present. 
Then, the periodic boundary condition along the $c$ axis is introduced 
instead, and $N (\ge 4,\ N={\it even})$ junctions are stacked in order to 
take non-uniformity into account. 
In such a condition plasma velocity of the stationary state automatically 
coincides with that of light in IJJs ($N$ is effectively infinite), though 
vanishing amplitude of electromagnetic wave (EMW) on surfaces 
is totally neglected and emitted EMW is parallel to layers. 
Even if a thousand of junctions are stacked, thickness of IJJ, $L_{z}$, 
is much smaller than the wavelength of emitted EMW, $\lambda$. 
In such a case emission originates from a point-like source (after 
two-dimensional reduction, by a line-like source in original 
three dimensions) and weaker than that parallel to layers. 
Such effect is approximately included as 
\begin{eqnarray}
\partial_{x'}\psi_{l}=B'_{\rm ext}+\tilde{B}'_{l}&,\ &
\partial_{t'}\psi_{l}=\langle E'_{l}\rangle+\tilde{E}'_{l},\\
\tilde{E}'_{l}=\mp Z \tilde{B}'_{l}&,\ &Z=z\sqrt{\epsilon'_{\rm c}/\epsilon'_{\rm d}},
\end{eqnarray}
with the dielectric constants of IJJs $\epsilon'_{\rm c}$ and of dielectrics 
$\epsilon'_{\rm d}$, respectively. Here dynamical parts of scaled boundary 
magnetic field $\tilde{B}'_{l}$ and electric field $\tilde{E}'_{l}$ are related 
with each other by $z \approx \lambda/L_{z}$,~\cite{Koshelev08a} though 
this naive evaluation of $z$ might be modified by excess magnetic fields 
from vertical directions omitted in two-dimnsional modeling.~\cite{Tachiki09} 

Width of the junction $L_{x}=86\mu$m, which is comparable 
to experimental scale and $B_{y}=0.02$T corresponds to 
``one Josephson vortex (JV) per layer" for this width, is divided 
into $80$ grids with the RADAU5 ODE solver.~\cite{radau5}
\medskip
\par
{\it Numerical results for $N=4$.} 
First, results for the minimum number of junctions are 
shown for various values of the surface impedance $Z$. 
In conventional picture of EMW emission from IJJs in in-plane magnetic 
fields, emission is driven by moving JVs, and the maximum emission is 
obtained when the distance between JVs coincides with the wavelength 
of EMW in a junction, which holds for $Z=1$.~\cite{Nonomura08}
However, this picture is the case only for in-plane magnetic fields 
applied to the in-phase state, which is stable only for small $Z$ 
without external magnetic fields.~\cite{Nonomura09} 
For $Z \ge 3$, emission from the cavity-resonance mode becomes 
stronger than that from the driven-JVs mode, and the fundamental 
mode becomes stronger than higher-harmonic modes. Then, only 
the fundamental mode is considered hereafter.
\begin{figure}
\includegraphics[height=6.0cm]{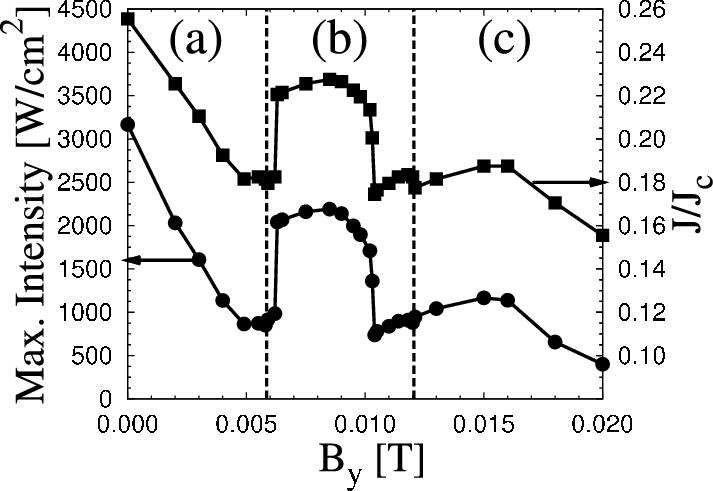}
\caption{\label{fig1}Field dependence of maximum intensity 
in the $n=1$ cavity resonance mode for $N=4$ (circles) 
at the optimal bias current (squares). Dips of the curves 
stand for the dynamical phase transitions between the 
(a) $\pi$-phase-kink ($\pi$-PK), (b) incommensurate-phase-kink 
(IPK), and (c) in-phase states.}
\end{figure}
\begin{figure}
\includegraphics[height=5.5cm]{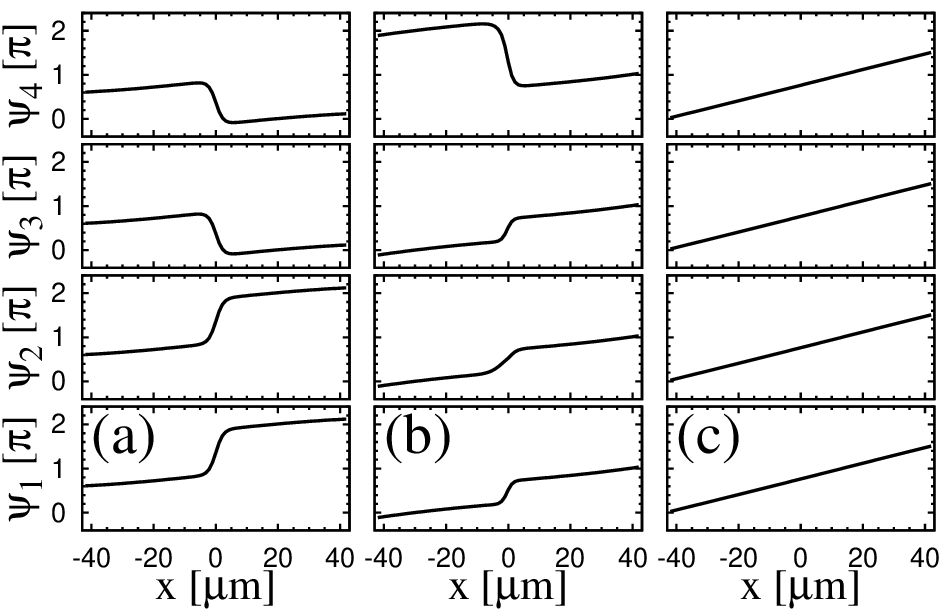}
\caption{\label{fig2}Snapshots of gauge-invariant phase differences 
in all junctions for $N=4$ at (a) $B_{y}=0.003$T, $J=0.210 J_{\rm c}$ 
(in the $\pi$-PK state; two $\pm\pi$ kinks), 
(b) $B_{y}=0.009$T, $J=0.226 J_{\rm c}$ (in the IPK state; 
three $+(1/2)\pi$ kinks and one $-(3/2)\pi$ kink), and 
(c) $B_{y}=0.015$T, $J=0.187 J_{\rm c}$ (in the in-phase state).}
\end{figure}
\smallskip
\par
\noindent
{\it (I) detailed results for $Z=30$.} 
As a typical value of the surface impedance, $Z=30$ is chosen.  
Maximum emission intensity and corresponding bias current 
are plotted versus in-plane magnetic field 
in Fig.\ \ref{fig1}. Procedure to obtain this figure is as follows: 
(1) to fix the in-plane magnetic field, (2) to sweep the 
bias current and evaluate stationary emission intensity, and 
(3) to draw the optimal set of stationary values. Two dips in this 
figure divide the three dynamical states characterized by the structure 
of gauge-invariant phase difference as displayed in Fig.\ \ref{fig2}: 
(a) the $\pi$-phase-kink ($\pi$-PK) state stable without external 
magnetic fields, (b) the incommensurate-phase-kink (IPK) state in the 
intermediate field region, and (c) the in-phase state. 
Field dependence of the current deriving the maximum intensity 
(optimal current) is also plotted in Fig.\ \ref{fig1} by squares. 
Its shape is quite similar to that of the maximum intensity, 
and this field profile will not be shown hereafter. 
The maximum intensity jumps up and down almost 
discontinuously inside (not on the boundaries) 
of the IPK state, which will be discussed later.
\begin{figure}
\includegraphics[height=11.5cm]{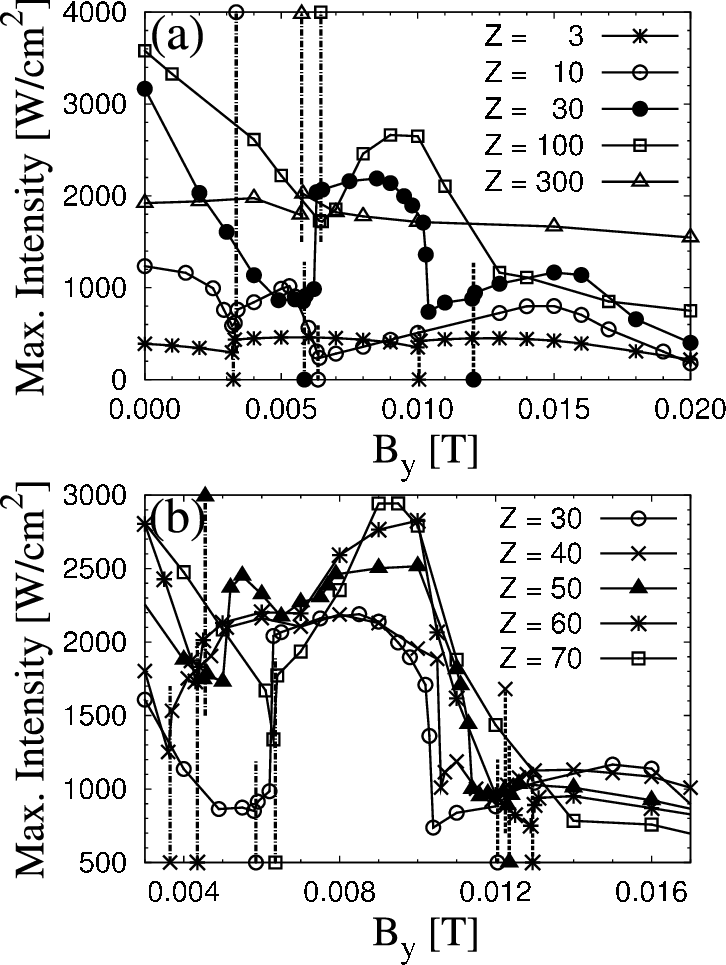}
\caption{\label{fig3}Field dependence of maximum intensity for $N=4$ 
for various values of the surface impedance $Z$: (a) for $Z=3$ (stars), 
$10$ (open circles), $30$ (full circles), $100$ (open squares) and $300$ 
(open triangles); (b) expanded figure for $Z=30$ (open circles), $40$ 
(saltires), $50$ (full triangles), $60$ (stars) and $70$ (open squares).
In each figure dash-dotted and dashed lines denote boundaries 
between the $\pi$-PK and IPK states and between the IPK and 
in-phase states, respectively.}
\end{figure}
\smallskip
\par
\noindent
{\it (II) results for other $Z$.}
Field dependence of 
maximum intensity for $N=4$ is given in Fig.\ \ref{fig3}(a) for 
$Z=3$, $10$, $30$, $100$ and $300$ together with transition fields 
for each $Z$ (dash-dotted lines between the $\pi$-PK and 
IPK states, and dashed lines between the IPK and in-phase 
states). This figure clearly indicates strong $Z$ dependence 
of the field profile of maximum intensity. 
For $Z=10$ and $30$, the first emission peak appears 
between the two ($\pi$-PK--IPK and IPK--in-phase) 
transition fields, and there exists the second peak 
at $B_{y}\approx 0.015 {\rm T}$ in the in-phase state. 
For $Z=100$ and $300$, on the other hand, 
the IPK--in-phase transition field does not 
locate for $B_{y}\leq 0.02 {\rm T}$, 
and no emission peak exists at $B_{y}\approx 0.015 {\rm T}$. 
These facts suggest that there would be a crossover of the field 
profile of maximum intensity between $Z=30$ and $100$.

Then, $Z$ dependence of the field profile of maximum 
intensity is displayed for $Z=30$, $40$, $50$, $60$ and $70$ in 
Fig.~\ref{fig3}(b) together with the transition fields for each $Z$. 
Although values of the $\pi$-PK--IPK transition field are close 
for $Z=30$ and $70$, this transition field is independent of the 
emission peak for $Z=30$, while it is linked with the peak for 
$Z=70$. This behavior changes between $Z=50$ and $60$. 
There exists similar change in behavior of the IPK--in-phase 
transition field between $Z=50$ and $60$. In addition, the 
characteristic peak at $B_{y}\approx 0.015 {\rm T}$ is visible 
up to $Z=40$, and it almost disappears for $Z\geq 50$. 
To summarize, there seems to exist a crossover in the 
field profile of maximum intensity at $Z \approx 50$.
\begin{figure}
\includegraphics[height=11.5cm]{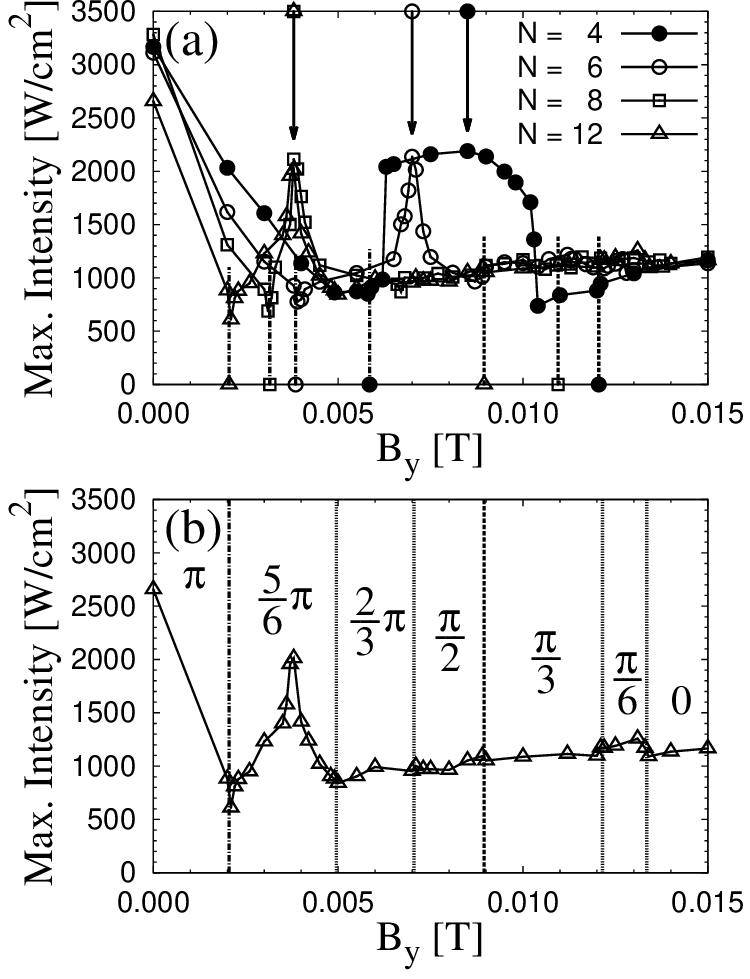}
\caption{\label{fig4}(a) Field dependence of maximum intensity for 
$Z=30$ for various number of junctions $N$: for $N=4$ (full circles), 
$6$ (open circles), $8$ (open squares) and $12$ (open triangles). 
Dash-dotted and dashed lines denote boundaries between the 
$\pi$-PK and the next IPK states and between the $\pi/2$-PK 
and the next IPK states, respectively. Arrows stand for the 
first peaks of the maximum intensity. 
(b) Field dependence of the data for $Z=30$ and $N=12$ 
is extracted in order to display various IPK states explicitly 
(``$0$" means the in-phase state).}
\end{figure}
\medskip
\par
{\it Numerical results for various number of junctions.} 
In order to compare the above results with experiments, 
we should clarify dependence of numerical results on the 
number of junctions $N$, and distinguish general properties 
independent of $N$ from special properties only for small $N$. 
As a typical value of the surface impedance $Z=30$ is chosen 
again, and  field dependence of maximum intensity 
for $N=4$ (full circles), $6$ (open circles), $8$ (open squares) 
and $12$ (open triangles) is displayed in Fig.~\ref{fig4}(a) 
together with the onset fields of the $\pi$-PK state 
(dash-dotted lines) and the $\pi/2$-PK state (dashed lines) 
and the first emission peaks (arrows). In order to show 
examples of various IPK states for larger $N$ explicitly, the data 
for $N=12$ are extracted in Fig.~\ref{fig4}(b), where 
the regions ``$\pi$", ``$\frac{5}{6}\pi$", ``$\frac{2}{3}\pi$", 
``$\frac{\pi}{2}$", ``$\frac{\pi}{3}$", ``$\frac{\pi}{6}$" and 
``$0$" represent the 
$[+\pi$-kink$\times 6 - \pi$-kink$\times 6 ]$, 
$[+\frac{5}{6}\pi$-kink$\times 7 - \frac{7}{6}\pi$-kink$\times 5 ]$, 
$[+\frac{2}{3}\pi$-kink$\times 8 - \frac{4}{3}\pi$-kink$\times 4 ]$, 
$[+\frac{\pi}{2}$-kink$\times 9 - \frac{3}{2}\pi$-kink$\times 3 ]$, 
$[+\frac{\pi}{3}$-kink$\times 10 - \frac{5}{3}\pi$-kink$\times 2 ]$, 
$[+\frac{\pi}{6}$-kink$\times 11 - \frac{11}{6}\pi$-kink$\times 1 ]$ 
and in-phase states, respectively. All possible combinations of 
IPK states appear as the in-plane magnetic field increases, and 
small jump or dip of the maximum intensity is observed at each 
transition field. 
\begin{figure}
\includegraphics[height=6.0cm]{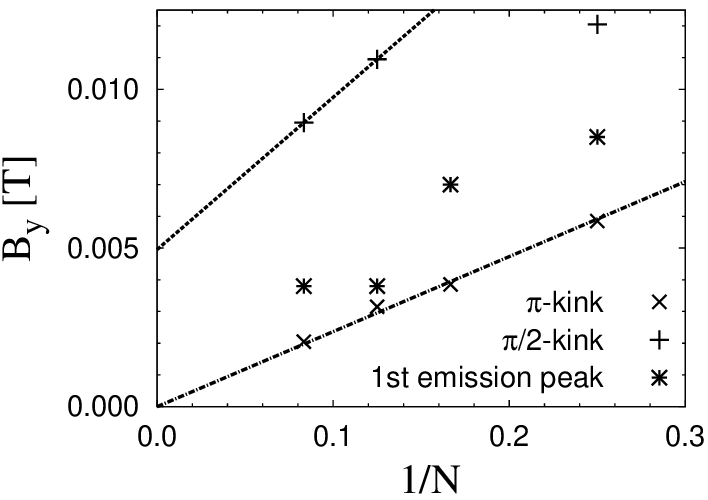}
\caption{\label{fig5}
Junction-number dependence of onset fields of the $\pi$-PK 
state (satires) and the $\pi/2$-PK state (crosses), and of fields 
at the first emission peak (stars) for $Z=30$. Dash-dotted and 
dashed lines are drawn as guide for eyes.}
\end{figure}

Finally, the onset fields of the $\pi$-PK and $\pi/2$-PK states and the 
field at the first emission peak are plotted versus $1/N$ in Fig.~\ref{fig5}. 
These data strongly suggest that the onset field of the $\pi$-PK state 
seems to converge to $B_{y}=0$ in the $N \to \infty$ limit (dash-dotted 
line). When these four data points are fitted by $B_{y}(N)=a/N+b$, 
we have $b=0.0002 \pm 0.0002$ [T]. 
On the other hand, the onset field of the $\pi/2$-PK states 
converges to a nonvanishing value (dashed line) with the 
same assumption. Similar preliminary results can be obtained 
for other IPK states, though the number of data is still limited. 
Moreover, the in-plane fields at the first emission peak coincide 
with each other for $N=8$ and $12$, and the extrapolated value 
is expected to be nonvanishing for large $N$.
\medskip
\par
{\it Discussions.}
Even if the periodic boundary condition is applied along the $c$ axis, 
dependence on the number of junctions $N$ is still not negligible 
as shown above. Since number of junctions in experiments is of 
order of $N \approx 1000$, extrapolation to large $N$ is crucial 
for reproducing experiments. On the other hand, maximum 
number of treatable $N$ is still not so large. CPU time is 
proportional to $N^{3}$ in this calculation, and relaxation time to 
reach stationary states rapidly increases as $N$ or $Z$ increases. 
For example, $t=5 \times 10^{3} \omega_{\rm p}^{-1}$ is long enough 
for $Z=10$ and $N=12$, but $t=2 \times 10^{4} \omega_{\rm p}^{-1}$ 
is at least necessary for $Z=30$ and $N=12$. Therefore, calculation 
for $N \ge 16$ or $Z=70$ in the same scale is difficult at present.

Then, alternative approach to numerical data is required, and the data 
for $N=4$ around the crossover region given in Fig.~\ref{fig3}(b) are 
reconsidered. The onsets of the first emission peak for $Z \le 50$ 
do not coincide with the boundaries of the IPK state including 
this peak, which suggests that the first emission peak is not 
related with the dynamical phase transitions in this parameter 
region. On the other hand, the onsets and the dynamical 
phase transitions seem to synchronize for $Z \ge 60$. 
After extrapolating to the large-$N$ limit, the zero-field 
peak and the dip at the onset of the $\pi$-PK state might 
cancel with each other and the $N$-independent first 
emission peak at $B_{y} \approx 0.003$T and the broad 
second emission peak at $B_{y} \approx 0.015$T in the 
in-phase state might remain for $Z \le 40$, while for $Z \ge 60$ 
both the onset of the $\pi$-PK state and the first emission peak 
shrink to zero field, and merely monotonically-decreasing 
field dependence of the higher-field side of the first 
emission peak remains.

The picture that a crossover between two different types 
of emission takes place at $B_{y} \approx 50$ can also 
be justified by the shape of emission peaks for $N=4$ by 
itself. The peak has sharp edges for $Z=30$, broadens 
to lower fields for $Z=40$, and splits for $Z=50$. 
The higher-field peak grows while lower-field one shrinks 
for $Z=60$, and the lower-field peak vanishes for $Z=70$.

These descriptions for large-$N$ behaviors of the present modeling 
seem consistent with the preceding experiments, namely the one by 
Yamaki {\it et al.}~\cite{Yamaki} with the case for $Z \le 40$, and the 
one by Welp {\it et al.}~\cite{Welp} with the case for $Z \ge 60$. 
If they are the case, the discrepancy between these two experiments 
is not serious contradiction but consequence of a little difference 
of experimental conditions to vary the surface impedance in the 
vicinity of the crossover point $Z \approx 50$. This argument 
also suggests that values of $Z$ in experiments may be 
much smaller than that naively expected from the relation 
$z \approx \lambda / L_{z}$ or $Z \approx 500$. It means 
that the argument by Tachiki {\it et al.}~\cite{Tachiki09} that 
$Z$ is effectively reduced by magnetic fields from vertical 
directions may be partially true, but that such effect is not 
so strong as they considered; they argued that $Z$ may be 
reduced so much that the in-phase emission is observed. 
Further studies to handle much larger numbers of 
junctions numerically or to manipulate the surface 
impedance experimentally are still necessary.

Finally, physical meaning of the IPK states is considered. 
When the in-plane magnetic field is applied, IPK between 
$0$ and $+\pi$ coupled with the in-plane field could be identified 
with ``$+\pi$-PK", and IPK between $-2\pi$ and $-\pi$ coupled 
with the in-plane field with ``$-\pi$-PK".  From this point of view, 
unbalance of ``$+\pi$-PKs" and ``$-\pi$-PKs" represents 
nonvanishing in-plane fields, and the in-phase state would 
be stable for $B_{y} \ge 0.01$T ($0.5$ JV per layer). 
The extrapolated onset field $B_{y} \approx 0.005$T of the 
$\pi/2$-PK state (Fig.~\ref{fig5}) is consistent with this picture.
\medskip
\par
{\it Summary.}
THz wave emission from intrinsic Josephson junctions 
in in-plane magnetic fields is investigated numerically. 
Cavity-resonant emission is observed for $Z \ge 3$ 
similarly to the case without external fields, and the 
$n=1$ emission mode is the strongest. 
There occur field-induced dynamical phase 
transitions between the $\pi$-phase-kink ($\pi$-PK) state, various 
incommensurate-phase-kink (IPK) states and in-phase state as the 
in-plane magnetic field increases. Investigation on dependence 
of physical quantities on junction numbers $N$ suggests that the 
$\pi$-PK state may be stable only without external magnetic fields 
for large $N$. In the field profile of maximum intensity there exist 
a crossover between two characteristic peaks at $\sim 0.15$ 
and $\sim 0.75$ JVs per layer~\cite{Yamaki} for $Z \le 40$ 
and monotonic decrease~\cite{Welp} for $Z \ge 60$, 
which suggests that controversy of the field profile in recent 
experiments~\cite{Welp,Yamaki} may be explained by slight 
difference in the surface impedance due to experimental conditions.
\medskip
\par
{\it Acknowledgments.}
The present work was partially supported by Grant-in-Aids for Scientific 
Research (C) No.\ 20510121 from JSPS.


\begin{thebibliography}{99}
\bibitem{Bae}
M.-H.~Bae {\it et al.}, Phys.\ Rev.\ Lett.\ {\bf 98}, 027002 (2007).
\bibitem{Ozyuzer}
L.~Ozyuzer {\it et al.}, Science {\bf 318}, 1291 (2007); 
see also K.~Lee {\it et al.}, Phys.\ Rev.\ B {\bf 61}, 3616 (2000).
\bibitem{Kadowaki}
K.~Kadowaki {\it et al.}, Physica C {\bf 468}, 634 (2008).
\bibitem{Sakai}
S.~Sakai {\it et al.}, J.\ Appl.\ Phys.\ {\bf 73}, 2411 (1993); 
S.~Sakai {\it et al.}, Phys.\ Rev.\ B {\bf 50}, 12905 (1994).
\bibitem{Matsumoto}
H.~Matsumoto {\it et al.}, Physica C {\bf 468}, 654, 1899 (2008).
\bibitem{Koyama}
T.~Koyama {\it et al.}, Phys.\ Rev.\ B {\bf 79}, 104522 (2009).
\bibitem{Lin08}
S.~Lin and X.~Hu, Phys.\ Rev.\ Lett.\ {\bf 100}, 247006 (2008).
\bibitem{Koshelev}
A.~E.~Koshelev, Phys.\ Rev.\ B {\bf 78}, 174509 (2008).
\bibitem{Nonomura09}
Y.~Nonomura, Phys.\ Rev.\ B {\bf 80}, 140506(R) (2009).
\bibitem{Welp}
U.~Welp {\it et al.}, on the APS March Meeting 2009, D34-1.
\bibitem{Yamaki}
K.~Yamaki {\it et al.}, Physica C {\bf 470}, S804 (2010).
\bibitem{Tachiki}
M.~Tachiki {\it et al.}, Phys.\ Rev.\ B {\bf 71}, 134515 (2005).
\bibitem{Koshelev08a}
A.~E.~Koshelev and L.~N.~Bulaevskii, Phys.\ Rev. B {\bf 77}, 014530 (2008).
\bibitem{Tachiki09}
M.~Tachiki {\it et al.}, Phys.\ Rev.\ Lett.\ {\bf 102}, 127002 (2009).
\bibitem{radau5}
http://www.unige.ch/$\tilde{\ }$hairer/software.html
\bibitem{Nonomura08}
Y.~Nonomura, J.\ Phys.: Conf.\ Ser. {\bf 150}, 052191 (2009).
\end{thebibliography}
\end{document}